\documentclass[prd,showpacs,11pt,nofootinbib]{revtex4-1}
\usepackage{color}
\usepackage[latin1]{inputenc}
\usepackage{amsmath}
\usepackage{amssymb}

\newcommand{\be}{\begin{equation}}
\newcommand{\ee}{\end{equation}}
\newcommand{\beann}{\begin{eqnarray*}}
\newcommand{\eeann}{\end{eqnarray*}}
\newcommand{\bea}{\begin{eqnarray}}
\newcommand{\eea}{\end{eqnarray}}

\newcommand{\bdm}{\begin{displaymath}}
\newcommand{\edm}{\end{displaymath}}

\usepackage[dvips]{graphicx}

\usepackage{epsfig}
\usepackage{graphicx}

\begin{document}
 
\title{Scalar models for the unification of the dark sector}

\author{Mahamadou H. Daouda}
\email{daoudah8@yahoo.fr}
\author{J\'{u}lio C. Fabris}
\email{fabris@pq.cnpq.br}
\author{Oliver F. Piattella}
\email{oliver.piattella@ufes.br}
\affiliation{Departamento de F\'isica, Universidade Federal do Esp\'irito Santo, avenida Ferrari 514, 29075-910 Vit\'oria, Esp\'irito Santo, Brazil}


\begin{abstract}
We review the difficulties of the generalized Chaplygin gas model to fit observational data, due to the tension between background and perturbative tests. We argue that such issues may be circumvented by means of a self-interacting scalar field representation of the model. However, this proposal seems to be successful only if the self-interacting scalar field has a non-canonical form. The latter can be implemented in Rastall's theory of gravity.
\end{abstract}


\maketitle

The dynamics of local structures in the universe cannot be explained in the framework of General Relativity by assuming the existence of only ordinary matter, composed essentially by baryons: a large amount of invisible matter with zero effective pressure is also necessary. This unknown form of matter, that escaped direct detection until now, is called {\it dark matter} and its existence is also required in order to explain the structure formation process in an expanding universe. The principal candidates for the role of dark matter are axions (light scalar particles predicted by the Grand Unified Theory) and neutralinos (massive, stable, supersymmetric particles) \cite{bertone}. Both these candidates exhibit a very small cross section, and do not emit any kind of electromagnetic radiation, thus possibly explaining the absence of a direct detection. On the other hand, these candidates are predicted on the basis of theories which have no experimental confirmation. This implies that the search for other possible explanations is necessary.
\par
The {\it dark side of the universe} contains another exotic component, named {\it dark energy}. The latter is necessary to explain the present stage of accelerated expansion of the universe, indicated by the analysis of type Ia Supernovae (SNIa) as standard candles \cite{caldwell}. It must be pointed out that there is a lot of discussion concerning the real status of SNIa as standard candles, but, even if such hypothesis is doubtful, the position of the first peak in the angular spectrum of the anisotropies in the cosmic microwave background radiation (CMBR) indicates a flat universe \cite{komatsu}. Therefore, since dark matter and baryons represent about the $30\%$ of the the critical density, the remaining $70\%$ must be in the form of another dark component, which, moreover, do not agglomerate via gravitational collapse. The latter property requires a negative pressure, which is also the condition necessary to drive the cosmic accelerated expansion. Such feature can be achieved by introducing a cosmological constant, re-interpreted as the quantum vacuum energy density, which is an inevitable consequence of quantum field theory in curved space-time. However, such interesting proposal is spoiled by a huge discrepancy between the predicted theoretical value and the one derived from observation. In view of this, many other candidates appeared (e.g. quintessence and $k$-essence models) all of them with some impressive successes and deceptive drawbacks \cite{padma}.
\par                                                                                                                                                                                                                                                                                                                                                                                                                                                                                                                                                                                                                                                            
Among the alternatives evoked to describe the dark sector of the universe, one interesting proposal are the so-called unified models. The prototype of these models is the Chaplygin gas (CG) \cite{moschella}, which is inspired by string theory: it is a fluid able to behave as dark matter in the past and as dark energy in the future, thus properly reproducing the expansion history of the universe. However, the CG suffers from some difficulties when it is tested against observation. This led to a more general formulation, the so-called generalized Chaplygin gas (GCG) \cite{berto1, neven}, which possesses a new parameter $\alpha$ ($\alpha = 1$ recovers the original CG). Unfortunately, a tension appears between the predictions of the model and various observational tests: the background ones (e.g. SNIa, BAO, H) favor
$\alpha < 0$ whereas the perturbative ones require $\alpha > 0$, in order to have a positive speed of sound \cite{colistete, berto2, finelli, piattella1}.
\par
In order to cope with this problem a possibility is to abandon the fluid description for the GCG, using instead a self-interacting scalar field. Unfortunately, this cannot be done using a canonical scalar field. We show that an interesting way out is to use a non-canonical self-interacting scalar field as suggested by Rastall's theory of gravity, an alternative
to General Relativity where the imposition of a strict conservation of the energy-momentum tensor is relaxed \cite{rastall}.
\par
Consider the GCG equation of state:
\begin{equation}
\label{eos}
p = - \frac{A}{\rho^\alpha}\;,
\end{equation}
where $\rho$ is the density and $\alpha$ and $A$ are positive parameters ($\alpha = 1$ corresponds to the original CG model). Inserting this expression in the continuity equation in a Friedmann-Lema\^{\i}tre-Robertson-Walker (FLRW) metric, i.e.
\begin{equation}
\dot\rho + 3\frac{\dot a}{a}(\rho + p) = 0\;,
\end{equation}
where $a$ is the scale factor and the dot indicates derivative with respect to the cosmic time,
one finds the following expression:
\begin{equation}
\rho = \rho_0\left[\bar{A} + (1 - \bar{A})a^{-3(1 + \alpha)}\right]^\frac{1}{1 + \alpha}\;,
\end{equation}
where $\bar{A} \equiv A/\rho_0^{1 + \alpha}$ and $\rho_0$ is the present-time GCG density (the present time is fixed by requiring $a_0 = 1$).
\par
The tension between background and perturbative constraints can be easily seen already from the equation of state (\ref{eos}): while in principle such definition allows for any real value of $\alpha$, at the perturbative level we have to deal with the square of the speed of sound, which has the following form:
\begin{equation}
c_s^2 \equiv \frac{d p}{d\rho} = \frac{\alpha A}{\rho^{1 + \alpha}}\;,
\end{equation}
which is positive only for $\alpha > 0$.
\par
The source of the tension comes from the fact that background tests favor negative values of $\alpha$ \cite{colistete}, which are excluded at perturbative level due to the occurrency of an imaginary speed of sound. The latter is due to the fluid representation adopted. It must be noted that such fluid representation emerges naturally from the Nambu-Goto action in the original CG formulation \cite{jackiw}, but the GCG has not the same clear connection, thus there is no fundamental reason to stay in the context of string theory. Therefore, it is possible to conceive a more fundamental representation for the GCG. A first possibility is to interpret the CGC as a self-interacting scalar field. On the other hand, a canonical self-interacting scalar field possesses a speed of sound equal to the speed of light, as we now prove.
\par
Following \cite{liddle}, we employ the formula
\begin{equation}\label{BDformula}
 \delta p = \hat{c_s}^2\delta\rho + 3aH\rho(1 + w)\left(\hat{c_s}^2 - c_a^2\right)\frac{\theta}{k^2}\;,
\end{equation}
which links pressure perturbations $\delta p$ to the energy density ones $\delta\rho$, both in a generic gauge, via $\hat{c_s}^2$, which is the speed of sound in the rest frame of the scalar field. In this formula, $c_a^2$ is the adiabatic speed of sound, defined as $c_a^2 \equiv \dot{p}/\dot{\rho}$; the equation of state parameter is $w \equiv p_\phi/\rho_\phi$; $k$ is the wavenumber coming from a normal mode decomposition and $\theta$ is defined via
\begin{equation}\label{theta}
 a(\rho + p)\theta \equiv \dot{\phi_0}\partial^i\delta\phi_i\;.
\end{equation}
A canonical self-interacting scalar field has the following expressions for the pressure and the density in the FLRW background:
\begin{equation}
\rho_\phi = \frac{\dot\phi^2}{2} + V(\phi)\;, \quad p_\phi = \frac{\dot\phi^2}{2} - V(\phi)\;,
\end{equation}
and it obeys the Klein-Gordon equation, which, in the FLRW background, reads
\begin{equation}
\ddot\phi + 3H\dot\phi = - V_\phi(\phi)\;.
\end{equation}
Hence, the adiabatic speed of sound for a canonical self-interacting scalar field has the form
\begin{equation}\label{adss}
c_a^2 = \frac{3H\dot{\phi_0} + 2V_\phi}{3H\dot{\phi_0}}\;.
\end{equation}
Substituting in Eq.~\eqref{BDformula} the expressions for $\delta\rho_\phi$ and $\delta p_\phi$, together with Eqs.~\eqref{theta} and \eqref{adss}, one obtains
\begin{equation}
 \hat{c_s}^2\left[\dot{\phi_0}\dot{\delta\phi} - \Phi\dot{\phi_0}^2 + V_{,\phi}\delta\phi + 3H\dot{\phi_0}\delta\phi\right]
= \left[\dot{\phi_0}\dot{\delta\phi} - \Phi\dot{\phi_0}^2 + V_{,\phi}\delta\phi + 3H\dot{\phi_0}\delta\phi\right]\;,
\end{equation}
which requires $\hat{c_s}^2 = 1$. This proves that a canonical scalar field cannot play the role of dark matter.
\par
So, how a self-interacting scalar field could represent the dark matter component? We have to consider a non-canonical scalar field. We adopt Rastall's gravity theory \cite{rastall}, where the energy-momentum tensor $T^{\mu\nu}$ does not obey the usual conservation law. Instead, the divergence of $T^{\mu\nu}$ reads
\begin{equation}
{T^{\mu\nu}}_{;\mu} = \kappa R^{;\nu}\;,
\end{equation}
where $R$ is the Ricci scalar and $\kappa$ is a constant. The field equations in Rastall's theory read
\begin{eqnarray}
R_{\mu\nu} - \frac{1}{2}g_{\mu\nu}R &=& 8\pi G\left(T_{\mu\nu} - \frac{\gamma - 1}{2}g_{\mu\nu}T\right)\;,\\
{T^{\mu\nu}}_{;\mu} &=& \frac{\gamma - 1}{2}T^{;\nu}\;,
\end{eqnarray}
where $\gamma$ is a dimensionless constant connected to $\kappa$. When $\gamma = 1$, General Relativity theory is recovered. Using the canonical form for the energy-momentum tensor of a scalar field, i.e.
\begin{equation}
T_{\mu\nu} = \phi_{,\mu}\phi_{,\nu} - \frac{1}{2}g_{\mu\nu}\phi_{,\rho}\phi^{,\rho} + g_{\mu\nu}V(\phi)\;,
\end{equation}
we obtain the following coupled equations:
\begin{eqnarray}
\label{Ein00} R_{\mu\nu} - \frac{1}{2}g_{\mu\nu}R &=& \phi_{,\mu}\phi_{,\nu} - \frac{2 - \gamma}{2}g_{\mu\nu}\phi_{,\alpha}\phi^{,\alpha} + g_{\mu\nu}(3 - 2\gamma)V(\phi)\;,\\
\Box\phi + (3 - 2\gamma)V_{,\phi} &=& (1 - \gamma)\frac{\phi^{,\rho}\phi^{,\sigma}\phi_{;\rho;\sigma}}{\phi_{,\alpha}\phi^{,\alpha}}\;.
\end{eqnarray}
From Eq.~\eqref{Ein00}, the following effective energy-momentum tensor can be read off:
\begin{equation}
\label{efetivo}
T_{\mu\nu}^{eff} = \phi_{,\mu}\phi_{,\nu} - \frac{2 - \gamma}{2}g_{\mu\nu}\phi_{,\alpha}\phi^{,\alpha} + g_{\mu\nu}(3 - 2\gamma)V(\phi)\;,
\end{equation}
implying the following expressions for the energy density and pressure:
\begin{equation}
\rho_\phi^{eff} = \frac{\gamma}{2}\dot\phi^2 \quad, \quad p_\phi^{eff} = \frac{2 - \gamma}{2}\dot\phi^2 - (3 - 2\gamma)V(\phi)\;.
\end{equation}
Using this expression in Eq.~\eqref{BDformula} to evaluate the speed of sound, one finds
\begin{equation}
c_s^2 = \frac{\gamma - 2}{\gamma}\;.
\end{equation}
This implies a vanishing speed of sound for $\gamma = 2$. In this case, the non-canonical self-interacting scalar field based on Rastall's theory may represent dark matter. On the other side, from the non-perturbative point of view, it can represents dark energy by a suitable choice of the potential. We will explore this possibility in what follows.
\par
Let us consider a self-interacting scalar field, with the effective energy-momentum tensor (\ref{efetivo}), with $\gamma = 2$:
\begin{equation}
T_{\mu\nu}^{eff} = \phi_{,\mu}\phi_{,\nu} -  g_{\mu\nu}V(\phi)\;.
\end{equation}
Inserting the FLRW metric, we have the following density and pressure associated with this scalar field:
\begin{equation}
\rho_\phi = \dot\phi^2 - V(\phi)\;, \quad p_\phi = - V(\phi)\;.
\end{equation}
Let us suppose that this density and pressure reproduce the background behaviour of the GCG model. Hence, in this case, we have,
\begin{eqnarray}
\dot\phi(a) &=& \sqrt{3\Omega_{c0}}\sqrt{g(a)^{1/(1 + \alpha)} - \bar A g(a)^{-\alpha/(1 + \alpha)}}\;,\\
V(a) &=& 3\Omega_{c0}\bar A g(a)^{-\alpha/(1 + \alpha)}\;,
\end{eqnarray}
where $g(a) \equiv \bar A + (1 - \bar A)a^{-3(1 + \alpha)}$. Hence, in order to have a zero speed of sound, the scalar model must obey the following equations:
\begin{eqnarray}
 R_{\mu\nu} &-& \frac{1}{2}g_{\mu\nu}R = 8\pi GT_{\mu\nu} + \phi_{,\mu}\phi_{,\nu} + g_{\mu\nu}V(\phi)\;,\\
 \Box\phi &+& V_\phi + \frac{\phi^{,\rho}\phi^{,\sigma}\phi_{;\rho;\sigma}}{\phi_{,\alpha}\phi^{,\alpha}} = 0\;,
\end{eqnarray}
where, just for future convenience, we have made the redefinition $V(\phi) \rightarrow - V(\phi)$. 
\par
Let us inspect now the perturbative behaviour of this system, computing scalar perturbations in the density contrast.
The perturbed equations in the synchronous coordinate condition read \cite{thais}:
\begin{eqnarray}
\label{rastall1}
\ddot\delta &+& 2\frac{\dot a}{a}\dot\delta - \frac{3}{2}\frac{\Omega_0}{a^3}\delta =
\dot\phi\dot\Psi - V_\phi\Psi\;,\\
\label{rastall2}
2\ddot\Psi &+& 3\frac{\dot a}{a}\dot\Psi + \left(\frac{k^2}{a^2} + V_{\phi\phi}\right)\Psi = \dot\phi\dot\delta\;,
\end{eqnarray}
where $\Psi = \delta\phi$ and $\delta$ is the density contrast of the matter component. Using the scale factor as independent variable, the above system of equations take the following form:
\begin{eqnarray}
 \delta'' &+& \left[\frac{2}{a} + \frac{f'(a)}{f(a)}\right] \dot\delta - \frac{3}{2}\frac{\Omega_0}{a^3f^2(a)}\delta =
\phi'\Psi' - \frac{V_\phi}{f^2(a)}\Psi\;,\\
2\Psi'' &+& \left[\frac{3}{a} + 2\frac{f'(a)}{f(a)}\right]\Psi' + \left[\frac{k^2}{a^2f^2(a)} + \frac{V_{\phi\phi}}{f^2(a)}\right]\Psi = \phi'\delta'\;,
\end{eqnarray}
where $f(a) = \dot a = \sqrt{\Omega_{m0}a^{-1} + \Omega_c(a)a^2}$ and $\Omega_c(a) = \Omega_{c0}g(a)^{1/(1 + \alpha)}$.
\par
\begin{figure}[htbp]
\includegraphics[width=0.3\linewidth]{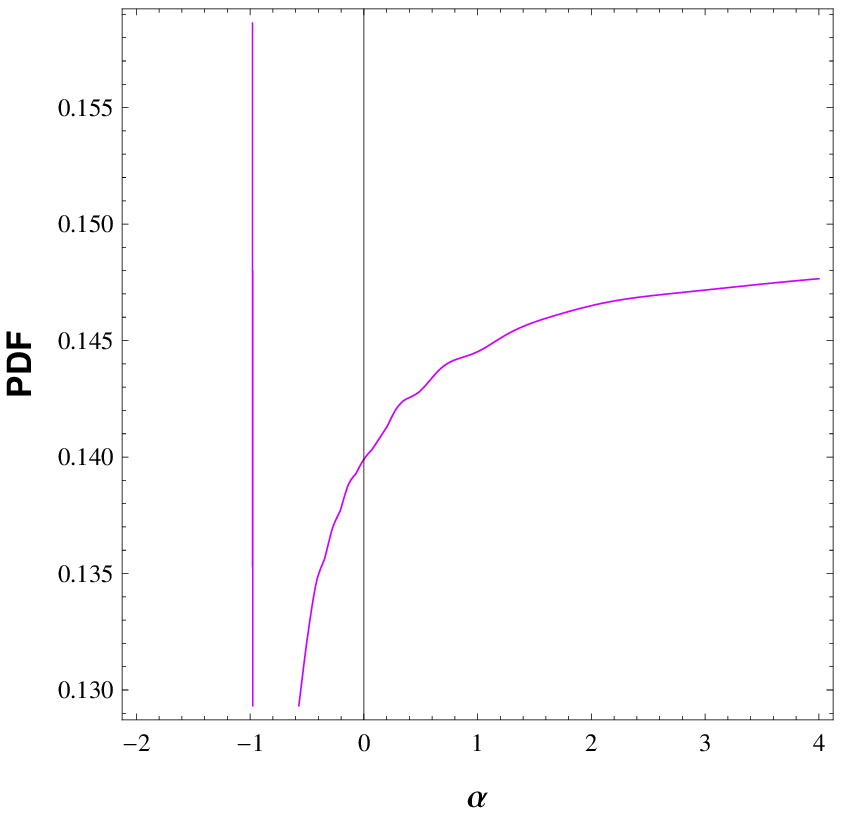}
\includegraphics[width=0.3\linewidth]{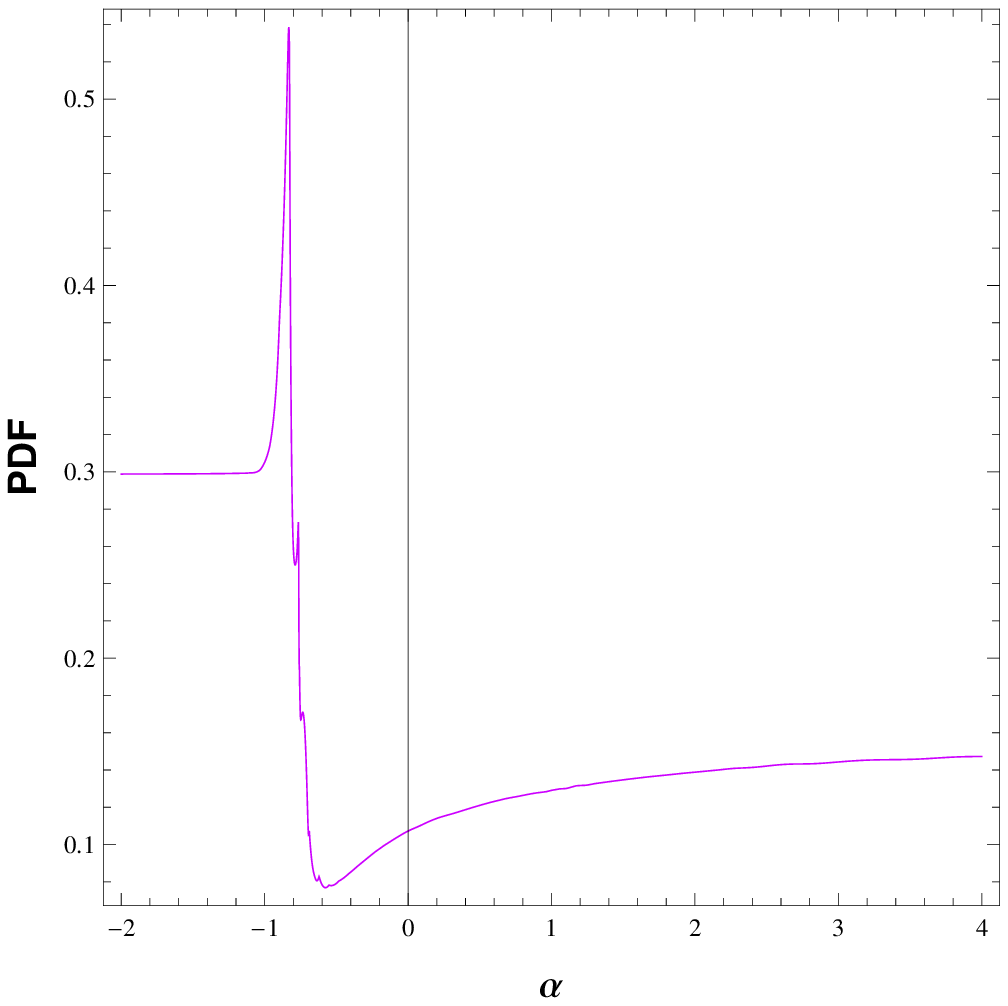}
\includegraphics[width=0.3\linewidth]{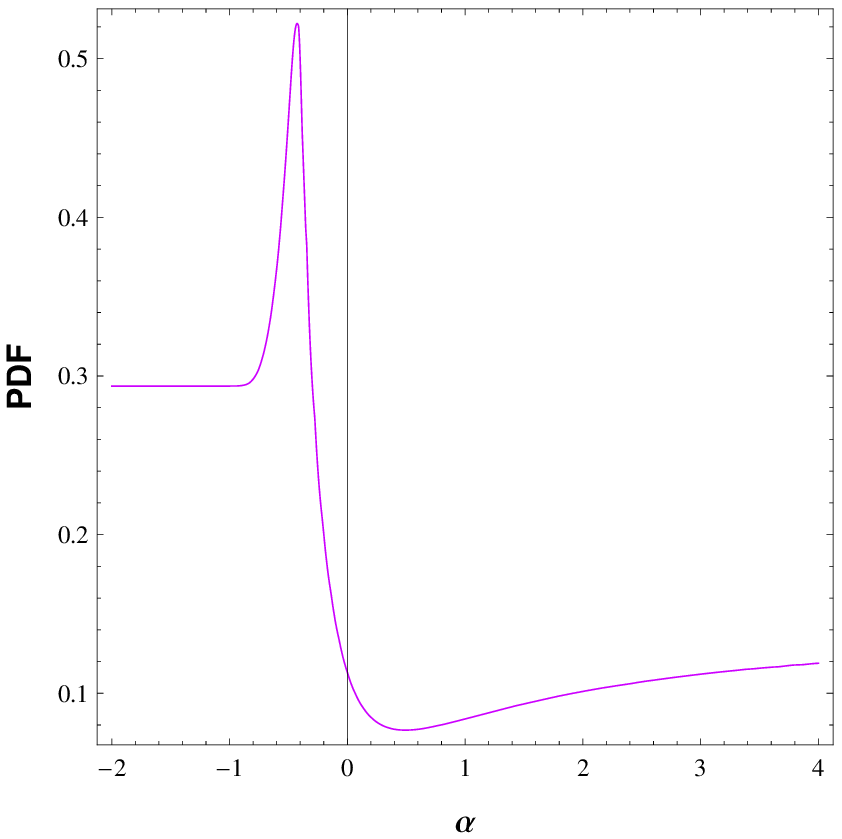}
\caption{One dimensional PDFs for the parameter $\alpha$ in the Rastall's scalar model with $\bar A = 0.1$; $\bar A = 0.5$ and $\bar A = 0.9$.}
\label{rastall-fig2}
\end{figure}

Using a Bayesian analysis and comparing the theoretical predictions of our model with the 2dFRGS data for the power spectrum of matter distribution in the universe, we find a significant probability region 
for $\alpha < 0$. The results are shown in figure \ref{rastall-fig2}, for some particular values of the parameter $\bar A$. For more details, see reference \cite{thais}.
In conclusion, if the GCG model is represented by a non-canonical scalar field, like the one suggested by Rastall's theory of gravity, 
the observational tension that plagues the GCG fluid model may disappear or, at least, be considerably alleviated. This fact may open new perspectives for the dark matter-dark energy unification program.

\section*{Acknowledgements}

We thank CNPq (Brazil) for partial financial support.


\end{document}